\documentclass[pra,twocolumn,superscriptaddress,showpacs,floatfix]{revtex4}

\usepackage{graphics}
\usepackage{graphicx}
\usepackage{bm}
\usepackage{amsmath}
\usepackage{amsfonts}
\usepackage{amssymb}
\usepackage{latexsym}
\usepackage{color}
\usepackage{bbold}
\usepackage{braket}

\begin{document}
	
\title{Quantum Jump Metrology}
\author{Lewis A. Clark}
\affiliation{The School of Physics and Astronomy, University of Leeds, Leeds LS2 9JT, United Kingdom}
\affiliation{Joint Quantum Centre Durham-Newcastle, School of Mathematics, Statistics and Physics, Newcastle University, Newcastle upon Tyne NE1 7RU, UK}
\author{Adam Stokes}
\affiliation{The School of Physics and Astronomy, University of Leeds, Leeds LS2 9JT, United Kingdom}
\affiliation{School of Physics and Astronomy, University of Manchester, Oxford Road, Manchester M13 9PL, United Kingdom}
\author{Almut Beige}
\affiliation{The School of Physics and Astronomy, University of Leeds, Leeds LS2 9JT, United Kingdom}

\date{\today}

\begin{abstract}
Quantum metrology exploits quantum correlations in specially prepared entangled or other non-classical states to perform measurements that exceed the standard quantum limit. Typically though, such states are hard to engineer, particularly when larger numbers of resources are desired. As an alternative, this paper aims to establish {\em quantum jump metrology}  which is based on generalised sequential measurements as a general design principle for quantum metrology and discusses how to exploit open quantum systems to obtain a quantum enhancement. By analysing a simple toy model, we illustrate that parameter-dependent quantum feedback can indeed be used to exceed the standard quantum limit without the need for complex state preparation.
\end{abstract}

\maketitle

\section{Introduction} \label{Introduction}

Over recent years, developing novel methods of conducting measurements with enhanced precision has become of increasing interest to a wide range of research areas and for a wide range of applications \cite{lifetime1,lifetime2,lifetime3,LIGO}. Classically, the best scaling in measurement uncertainty that one can achieve is the standard quantum limit. This limit applies when the variance $\left(\Delta \varphi \right)^2$ of a measured parameter $\varphi$ is inversely proportional to $N$,
\begin{eqnarray}
\left( \Delta \varphi \right)^2 & \propto & N^{-1} \, ,
\end{eqnarray}
where $N$ is called a {\em resource}. Depending on the type of measurement considered, $N$ could correspond to a number of different quantities. For instance, $N$ could be the number of photons involved in an interferometric phase shift measurement between two pathways of light. Alternatively, $N$ might relate to the total length of the measurement process. 

As is well known, it is possible to exploit the more counter-intuitive properties of quantum physics, like entanglement, to increase the precision of measurements and to overcome the standard quantum limit. 
This can be done in a variety of ways \cite{Caves,Squeeze,Berry2,ECS2,Gerry0,Q.Met,Q.Met2,Berry,M&M,Gerry,Gerry4,ECS,Kok,mm',Paul,Gerry3,Fluctuation}  (see for example Ref.~\cite{review} for a recent review). Using quantum resources, the variance $\left(\Delta \varphi \right)^2$ of a measured parameter $\varphi$ can be shown to possibly scale as 
\begin{eqnarray}
\left( \Delta \varphi \right)^2 & \propto & N^{-2} 
\end{eqnarray}
which is known as the Heisenberg limit \cite{braunstein,Kok-book}. In principle, even this limit can be overcome \cite{5} but this requires non-linear system dynamics which are difficult to induce. Even the preparation of highly entangled resources usually poses serious experimental challenges. This means that although very high scaling is achievable within the theoretical framework, it is not likely to be achieved on a large scale in a lab in the near future. This makes the development of alternative approaches for potentially immediate practical quantum technology applications desirable, even when these do not necessarily realise the full potential of the Heisenberg limit. 

In this paper, we have a closer look at quantum metrology schemes that do not rely on entanglement as a resource and are therefore easier to implement experimentally \cite{Braun,Openmet,PRA,Ref1,Ref2,Ref3,review2}. Such schemes exploit the strong temporal correlations that are known to exist in the system dynamics of open quantum systems with generalised sequential measurements \cite{Kok2,Kok3,Maybee}. Also, monitoring the environment has been shown previously to have benefits for quantum metrology even when using entanglement \cite{Albarelli}.  Although the modelling of such systems is well understood \cite{Molmer,Reset,Carmichael,Breuer-Petruccione,Wiseman-Milburn}, it is in general difficult to design and analyse these quantum metrology schemes, since the derivation of the scaling laws for quantum metrology schemes in closed systems do not automatically extend to open systems and require novel insight \cite{Braun,Openmet}. Some of the currently known results only apply to specific systems, which can be analysed analytically by drawing analogies to closed systems \cite{review2}. In other cases, scaling laws for measurement errors can only be obtained through extensive numerical simulations of the proposed measurements \cite{PRA}.

\begin{figure}[t] 
\centering
\includegraphics[width=0.48\textwidth]{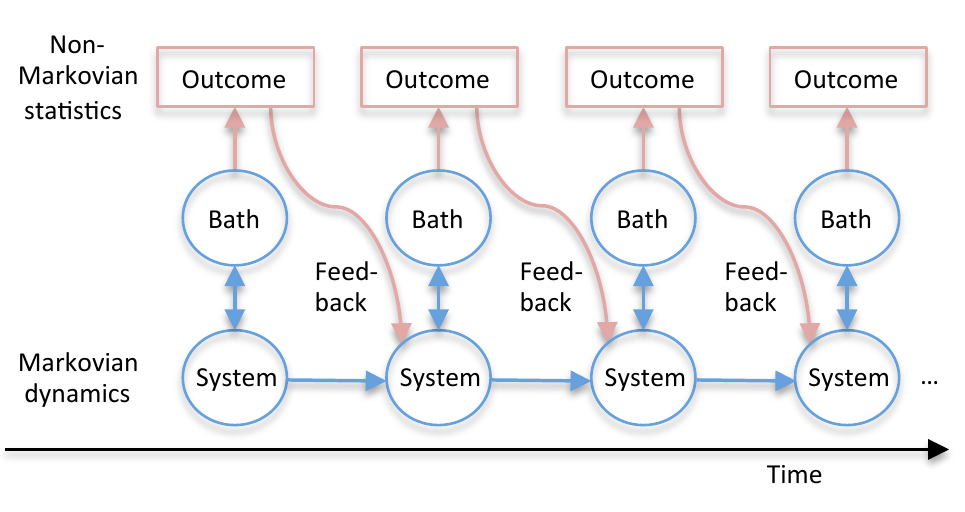}
\caption{[Colour online] Schematic view of the generation of non-Markovian measurement statistics via sequential measurements in open quantum systems. In every time step, an interaction between the quantum system and its surrounding bath results in the generation of a measurement signal. For instance, an open quantum system might emit a photon or not. Depending on the measurement result, the state of the quantum system is altered. Although the system dynamics are Markovian, the same does not need to be true for the generated measurement sequence.}
\label{generalfig}
\end{figure}

This paper aims to establish {\em quantum jump metrology} which is based on generalised sequential measurements as a general design principle for developing quantum metrology schemes. While it is already known that quantum-enhanced metrology does not require entanglement as a resource \cite{review2,Baumgartz}, our motivation here is to provide a straightforward methodology for obtaining such an enhancement without it. In the following, we obtain two necessary (although not sufficient) conditions for obtaining a quantum enhancement when measuring an unknown parameter $\varphi$. As we shall see below, quantum jump metrology exploits strong temporal correlations in the statistics of measurements which can be described by Kraus operators \cite{Kraus}. The first condition demands that at least some of these Kraus operators should depend on $\varphi$ in a non-trivial way. This could mean resetting the quantum system into a $\varphi$-dependent state when a certain measurement outcome is obtained, as illustrated in Fig.~\ref{generalfig}, which considers an open quantum system. Secondly, Kraus operators which correspond to different measurement outcomes should not commute with each other which is a necessary criteria for strong temporal correlations in the measurement statistics. 

Systems with long-range temporal correlations in their measurement statistics are well-known to be useful in classical computer science, and are classified as Hidden Markov Models. Their name derives from the fact that they progress randomly from one internal state to another, which remains unobserved (hidden), while producing a stochastic output sequence \cite{HMM1,HMM2,HMM3}. Although based on Markovian system dynamics, Hidden Markov Models can produce non-Markovian measurement sequences (c.f.~Fig.~\ref{generalfig}). The same applies to quantum versions of Hidden Markov Models, so-called Hidden Quantum Markov Models \cite{Wiesner,Monras,Schuld,Biamonte,Bristol}. A standard example for Hidden Quantum Markov Models are open quantum systems with quantum feedback \cite{HQMM}. In other words, this paper proposes to utilise Hidden Quantum Markov Models in quantum metrology. 

Recent research has shown that Hidden Quantum Markov Models are able to produce stronger temporal measurement correlations than their classical counter-parts, even when using significantly less resources to store information \cite{Monras,indian,Yeh}. Similarly, quantum neural networks have also been shown to offer an enhancement versus their classical counterpart \cite{QNN}.  The dynamics of these quantum versions exhibit a higher degree of complexity which can result in an extreme advantage for a wide range of computational tasks, like the simulation of complex systems \cite{Crutchfield}. 

There are five sections in this paper. In Sec.~\ref{PET} we give a brief overview of parameter estimation theory, especially Fisher information and scaling laws. In Sec.~\ref{QJM}, we demonstrate how parameter estimation theory can be applied to analyse temporal sequential measurements in open quantum systems. Here we introduce the basic idea of quantum jump metrology and illustrate it with two toy model cases. In Sec.~\ref{Toy}, we analyse a concrete physical system with more practical relevance and discuss how to exceed the standard quantum limit in open quantum systems. Finally we conclude and discuss the results of our work in Sec.~\ref{Conclusions}.

\section{Parameter Estimation Theory} \label{PET}

To see the benefit of quantum metrology, we shall consider a brief mathematical analysis of parameter estimation theory and give an overview of the Fisher information and Cram\'er-Rao bound (for more details see for example Ref.~\cite{Kok-book}).  The classical Fisher information is useful in determining the precision of an estimator ${\hat \varphi}({\bf x})$ of some parameter $\varphi$. The estimate ${\hat \varphi}({\bf x})$ is assumed to depend on the values of some data-string ${\bf x} \in {\mathbb R}^N$, for some $N \in {\mathbb N}$, of a real random vector ${\bf X}$ defined over a Kolmogorov probability space. The Fisher information associated with the probability density $P_\varphi$ is defined by
\begin{align}\label{F}
F(P_\varphi) & = \int d^N x P_\varphi({\bf x}) [\partial_\varphi \ln P_\varphi({\bf x})]^2 \nonumber \\
& = \int d^N x {[\partial_\varphi P_\varphi({\bf x})]^2 \over P_\varphi ({\bf x})} \, .
\end{align}
The Fisher information is additive for independent sources of knowledge; $ F(P)= F(P_1)+F(P_2)$ whenever $P(x_1,x_2) = P^1(x_1) P^2(x_2)$.

The Cram\'er-Rao bound gives a lower bound on the precision of an estimate ${\hat \varphi}$ using the Fisher information. The bound is
\begin{align}\label{CR}
\langle \Delta{\hat \varphi} ^2\rangle_{P_\varphi} \geq {1\over F(P_\varphi)}+ \langle \Delta{\hat \varphi}\rangle_{P_\varphi}^2 \geq {1\over F(P_\varphi)} \, ,
\end{align}
where for an unbiased estimate $\langle \Delta{\hat \varphi}\rangle_{P_\varphi}^2 =0$. The proof of Eq.~(\ref{CR}) involves a straightforward application of the Cauchy-Schwarz inequality $\|{x}\| \|{y}\| \geq |\langle x,y\rangle |^2$ applied to the natural inner-product defined over the $L^2({\mathbb R}^N)$ function space \cite{braunstein,Kok-book}.

In practice one gathers information about a physical system in the form of a list of numbers obtained by querying the system. The values $x_N$ can be viewed as the result of querying a physical system $N$ times, which would be equivalent to having an ensemble of $N$ identically prepared independent systems that have the same state ${\tilde P}_\varphi =P_\varphi(x_i)$, $\forall i=1,...,N$. More generally, systems that are independent but not necessarily identically prepared, are described by a {\em product} distribution $P_\varphi ({\bf x}) = \prod_{i=1}^N P^i_\varphi(x_i)$. For such a distribution the Cram\'er-Rao bound and the additivity of the Fisher information yield the bound
\begin{align}\label{CR2}
\langle \Delta{\hat \varphi} ^2\rangle_{P_\varphi} \geq {1\over N F^{\rm max}} \, ,
\end{align}
where $F^{\rm max} = \max_{x_i} F(P^i_\varphi(x_i))$. The number $N$ is called the {\em resource} and is what was discussed in the previous section in terms of the limits. It is the number of times one has queried the system to gather information in the form of a list of numbers ${\bf x}$. The bound from Eq.~(\ref{CR}) yields the so-called standard quantum limit scaling of $1/\sqrt{N}$ for the lower bound of $\sqrt{\langle \Delta{\hat \varphi} ^2\rangle_{P_\varphi}}$.

In quantum metrology one considers a quantum system whose density matrix $\rho_\varphi$ depends on an unknown parameter $\varphi$. According to quantum theory, a measurement of the physical system yields an outcome ${\bf x}$ with probability $P_\varphi({\bf x})={\rm tr}(E_{\bf x}\rho_\varphi)$, where $E_{\bf x}$ is a positive operator-valued measure (POVM) describing the measurement process. The quantum Fisher information can be defined as
\begin{align}\label{qF}
F_Q(\rho_\varphi) = \max_{E_{\bf x}} F(P_\varphi({\bf x})) \, .
\end{align}
The quantum Cram\'er-Rao bound
\begin{align}
\langle \Delta{\hat \varphi} ^2\rangle_{\rho_\varphi} \geq {1\over F_Q(\rho_\varphi)} 
\end{align}
then follows from the Cram\'er-Rao bound (\ref{CR}). The Quantum Fisher information is additive in that $F_Q(\rho^1_\varphi \otimes \rho^2_\varphi )= F_Q(\rho^1_\varphi)+F_Q(\rho^2_\varphi)$ whenever the composite state $\rho_\varphi=\rho_\varphi^1 \otimes\rho_\varphi^2$ varies with $\varphi$ according to $\partial_\varphi \rho_\varphi = i[\rho_\varphi,h^1_\varphi\otimes I^2 +I^1\otimes  h^2_\varphi]$ with $h_\varphi^{1,2}$ being a Hermitian operator. In this case, for an uncorrelated $N$-part state $\rho_\varphi = \bigotimes_{i=1}^N \rho_\varphi^i$ the quantum Cram\'er-Rao bound yields
\begin{align}\label{SQL}
\langle \Delta{\hat \varphi} ^2\rangle_{\rho_\varphi} \geq {1\over N F^{\rm max}_Q} \, ,
\end{align}
where $F_Q^{\rm max} = \max_i F_Q(\rho_\varphi^i)$. The above bound gives the standard quantum limit scaling for the precision. Since each system making up the $N$-part composite system is queried once in a measurement, the number $N$ coincides with the number of queries made.

One way to obtain an enhancement over the scaling of $1/N$ of the standard quantum limit in Eq.~(\ref{SQL}) is to consider an $N$-part system that is prepared in an entangled state. Since for an entangled state the Fisher information is not additive the bound in  Eq.~(\ref{SQL}) does not follow from the Cram\'er-Rao bound. It is then possible to improve upon the standard scaling to obtain the Heisenberg scaling $\langle \Delta{\hat \varphi} ^2\rangle_{\rho_\varphi} \sim 1/N^2$ \cite{braunstein,Kok-book}. The crucial ingredient in obtaining this enhancement is the breakdown of additivity of the quantum Fisher information due to the presence of correlations within the $N$-part system.  Then, although not always, it is possible that the Fisher information will scale greater than linearly.

\section{Quantum Jump Metrology} \label{QJM}

We will now introduce a method of creating non-additive Fisher information that can produce a non-linear scaling with respect to the resource without the need for entangled state preparation. Concrete examples of quantum metrology schemes, which do not require entanglement as a resource, can already be found in the literature \cite{Braun,Openmet,PRA,Ref1,Ref2,Ref3,review2}. In the following, we aim to establish a general design principle for quantum metrology schemes that are based on sequential measurements and quantum feedback and which we refer to as {\em quantum jump metrology}. By calculating the Fisher information for relatively simple two-level toy models, it is shown that quantum jump metrology schemes are indeed able to exceed the standard quantum limit. 

In contrast to metrology schemes that require entanglement as a resource and which are difficult to realise experimentally, quantum jump metrology schemes are easily scalable. As we shall see below, in order to obtain a quantum enhancement in the uncertainty scaling, all that is required is correlations. Entanglement is one special example of such correlations, but its presence is not a necessary criterion. To obtain more insight as to where the enhancement comes from, we present a thorough analysis of the Fisher information for specific examples. We discuss how to introduce the necessary quantum correlations in open quantum systems with quantum feedback and identify types of processes that can be useful for quantum metrology. We expect that our results can be used to guide the design of quantum metrology schemes in open quantum systems.

\subsection{Correlated distributions yield non-additive Fisher information}

Temporal quantum correlations \cite{Kok2,Kok3} and sequential measurements \cite{Ref1,Ref2,Ref3} in open quantum systems are known to constitute an interesting resource for quantum technology applications.  The analysis of the previous section shows that enhancement over the standard quantum limit can be obtained when additivity of the quantum Fisher information fails to hold. The quantum Fisher information is simply a specific type of classical Fisher information having the form of Eq.~(\ref{qF}). Of course one can consider the precision of parameter estimates without restricting one's attention to the quantum Fisher information, especially in the case we consider here where the measurement outcomes are effectively a classical string of data. The standard quantum limit scaling seen in Eq.~(\ref{CR2}) follows from the Cram\'er-Rao bound in Eq.~(\ref{CR}) when the Fisher information is additive, i.e., when the probability density $P_\varphi({\bf x})$ is uncorrelated; $P_\varphi({\bf x}) = \prod_{i=1}^N P_\varphi(x^i)$. When there are correlations present within $P_\varphi({\bf x})$ the standard quantum limit-scaling does not necessarily follow, which allows for the possibility of obtaining enhanced precision. One way to achieve such enhancement is to consider a distribution of the form $P_\varphi({\bf x};E_{\bf x})={\rm tr}(E_{\bf x}\rho_\varphi)$ in which $\rho_\varphi$ is an entangled quantum state and $E_{\bf x}$ is a POVM. However, this is by no means the only way to obtain a correlated distribution $P_\varphi({\bf x})$. The use of entanglement is hence not the only means by which to obtain enhanced precision.

\subsection{Producing temporal correlations}

In this section we consider a different approach. Our aim is to determine precision bounds on parameter estimates when the queries of a system are represented by parameter-dependent POVMs. Suppose generalised measurements are performed on a single qubit at short time intervals and the only possible measurement outcomes are 0 or 1. Moreover, we assume in the following, that these measurements trigger a parameter dependent back-action and describe their overall effect on the state of the single qubit by parameter-dependent Kraus operators $K_{0,1}(\varphi)$ \cite{Kraus}. They must satisfy the completeness relation
\begin{eqnarray} \label{Kraus-op}
\sum\limits_n K^\dagger_n K_n & = & \mathbb{1} \, .
\end{eqnarray}
Consequently, the measurement statistics created by the random dynamics of the single qubit exhibits so-called {\em quantum jumps} \cite{jump}. Any sequence of such quantum jumps corresponds to a concrete quantum trajectory within the single-qubit Hilbert space.

One way of implementing parameter-dependent Kraus operators in the dynamics of a two-level system is to create an interaction between the single qubit and an auxiliary system, a so-called ancilla, which is measured and reset after every discrete time step of the evolution. Since the ancilla always ends up eventually, i.e.~at the end of every measurement, in exactly the same state, the dynamics of the single qubit may be described by a sequence of Kraus operators. In the following, we have a closer look at possible measurement schemes that are capable of quantum-enhance precision. In the approach described here there is no need to prepare entangled states of the system measured.

Firstly, we shall demonstrate that the output produced by a sequence of generalised measurements can indeed be highly correlated. The distribution $P_\varphi({\bf x})$ of outcomes after $N$ sequential queries on the parameter $\varphi$ is given by
\begin{align}\label{d}
P_\varphi({\bf x}) = {\rm tr}(K_{x_N}K_{x_{N-1}}...K_{x_1}\rho K^\dagger_{x_1}...K^\dagger_{x_{N-1}}K^\dagger_{x_N}) \, ,
\end{align}
where $\rho$ is the initial state of the system and where $x_i=0,1$ is the outcome of the $i'{\rm th}$ measurement. In general the distribution $P_\varphi({\bf x})$ is correlated, i.e., is not of the product form $\prod_{i=1}^N{\tilde P}_\varphi(x^i)$. Even when the reduced dynamics of the quantum system are Markovian, the distribution $P_\varphi({\bf x})$ does not result from a Markov chain of outcome events \cite{Monras,HQMM}. To see this note that at each step $i=1,...,N$ the operators $K_{0,1}$ respectively select subensembles of systems for which outcomes $0,1$ were obtained. The complete ensemble at step $i$ is therefore represented by a density matrix
\begin{align}
\rho(i) &= {\cal T}(\rho(i-1)) \nonumber \\ &:= K_0\rho(i-1)K_0^\dagger+K_1\rho(i-1)K_1^\dagger \, ,
\end{align} 
where ${\cal T}$ denotes the Markovian evolution map that propagates the system's state to the next step. Consider the example $N=3$. In this case, we have
\begin{align}\label{c1}
P_\varphi(x_3|x_2,x_1) = {{\rm tr}(K_{x_3}K_{x_2}K_{x_1}\rho K^\dagger_{x_1}K^\dagger_{x_2}K^\dagger_{x_3}) \over {\rm tr}(K_{x_2}K_{x_1}\rho K^\dagger_{x_1}K^\dagger_{x_2})} \, ,
\end{align}
whereas
\begin{align}\label{c2}
P_\varphi(x_3|x_2) := {{\rm tr}(K_{x_3}K_{x_2}{\cal T}(\rho) K^\dagger_{x_2}K^\dagger_{x_3}) \over {\rm tr}(K_{x_2}{\cal T}(\rho) K^\dagger_{x_2})} \, .
\end{align}
In general the right-hand-side of Eq.~(\ref{c2}) is not equal to the right-hand-side of Eq.~(\ref{c1}), therefore the random variable sequence $X_1\to X_2 \to X_3$ is not a Markov chain. Since the state of the system after each measurement depends on the outcome obtained, the probability density $P_\varphi({\bf x})$ can become highly correlated throughout the course of the $N$-measurements.  This is the result of the system having coherences that are not necessarily destroyed in Kraus measurements. The presence of correlations in the distribution in Eq.~(\ref{d}) means that the standard quantum limit does not necessarily follow from the Cram\'er-Rao bound for the associated Fisher information.

\subsection{Implementations}

To illustrate the idea of determining precision bounds within the context described above we now consider some simple examples involving just a single qubit.  All of the examples can be implemented by applying simple operations to a qubit and an ancilla \cite{Kraus,Monras,HQMM}.  First we examine a system that does not produce an enhancement but produces the usual scaling of the standard quantum limit in order to show a simple example of how it may be calculated in this context. Afterwards, we discuss an example that does produce an enhanced scaling.

\subsubsection{An example without enhanced precision}

Our first example assumes a qubit system with the two Kraus operators 
\begin{align}\label{ch}
K_0 & = \left( {\begin{array}{cc}
 \cos (\varphi) & 0 \\
  0 & \cos (\varphi)
 \end{array} } \right), \nonumber \\  K_1 & = \left( {\begin{array}{cc}
 0 & \sin (\varphi) \\
  \sin (\varphi) & 0 \, .
 \end{array} } \right) 
\end{align}
These operators could be generated by taking an ancilla initially prepared in $\ket{0}$ and performing a Pauli operation $\sigma_x = \ket{1}\bra{0} + \ket{0} \bra{1}$ on the system qubit and ancilla with probability $\sin^2(\varphi)$.  By then measuring the ancilla in either state $\ket{0}$ or $\ket{1}$, we obtain the above Kraus operators. These satisfy the relations  
\begin{align}\label{prop}
K_{0,1}=K_{0,1}^\dagger, \qquad K_0^2+K_1^2 = \mathbb{1},
\end{align}
and
\begin{align}\label{prop2}
[K_0,K_1]=0 \, .
\end{align}
The first and second property ensure that the $K_x$ with $x=0,1$ are indeed Kraus operators. Since  $K_0=\cos (\varphi) \mathbb{1}$, $K_0$ and $K_1$ commute, which makes them amenable to analytic calculations. Moreover $K_1=\sin (\varphi) \, \sigma_x$ so that $K_0^2 = \cos^2 (\varphi) \mathbb{1}$ and $K_1^2 = \sin^2 (\varphi) \, \mathbb{1}$. For this choice of Kraus operators and for fixed $\varphi$ the number of different values of $P_\varphi({\bf x})$ is only $N$, because if ${\bf x}$ and ${\bf x}'$ contain the same number of zeros and ones then $P_\varphi({\bf x})=P_\varphi({\bf x}')$. Since ${\rm tr} (\rho) = 1$ for any initial state $\rho$, if ${\bf x}$ contains $k_x$ zeros, we get
\begin{align}
P_\varphi ({\bf x}) &= {\rm tr}(K_0^{2k_x} K_1^{2(N-k_x)}\rho)\nonumber \\ &=\cos^{2k_x} (\varphi) \, \sin^{2(N-k_x)} (\varphi) \, ,
\end{align}
where we have used the cyclicity of the trace and Eqs.~(\ref{prop}) and (\ref{prop2}). The ${\bf x}$ are binomially distributed in that the number of ${\bf x}$'s with $k_x$ zeros and $N-k_x$ ones is ${N\choose k_x}$. We can calculate the Fisher information (c.f.~Eq.~(\ref{F})) associated with $\rho_\varphi$ as
\begin{align}\label{f}
F &= \sum_{\bf x} {[\partial_\varphi P_\varphi({\bf x})]^2\over P_\varphi({\bf x})} \nonumber \\ &= \sum_{k_x=0}^N {N\choose k_x} (N-2k_x+N\cos (2\varphi))^2 \nonumber \\ & \quad \times  \cos^{2(k_x-1)} (\varphi) \sin^{2(N-k_x-1)} (\varphi)  \nonumber \\ &=4N \, .
\end{align}
Thus, for the choices in Eq.~(\ref{ch}) we get the standard quantum limit scaling from the Cram\'er-Rao bound in Eq.~(\ref{CR});
\begin{align}
\langle (\Delta \varphi)^2\rangle \geq {1\over 4N} \, .
\end{align}
This result is due to the nature of the distribution $P_\varphi({\bf x})$, which can in fact be written as a product distribution $\prod_{i=1}^N P_\varphi (x_i)$. To see this, note that in this particular example $P_\varphi^i(x_i) = P_\varphi^j(x_j)$ whenever $x_i=x_j$, so $P_\varphi^i$ is actually independent of $i$. Over all steps $i=1,\dots,N$ there are only two possible probabilities;
\begin{align}
&P_\varphi^i(0)=P_\varphi(0) = {\rm tr}(K_0^2\rho) = \cos^2 (\varphi) \, , \nonumber \\ &P_\varphi^i(1)=P_\varphi(1)={\rm tr}(K_1^2\rho) = \sin^2(\varphi) \, .
\end{align}
We therefore have
\begin{align}\label{pd}
P_\varphi({\bf x}) &= \cos^{2k_x} (\varphi) \sin^{2(N-k_x)} (\varphi) \nonumber \\ & = P_\varphi(0)^{k_x} P_\varphi(1)^{N-k_x} \nonumber \\ &= \prod_{i=1}^N P_\varphi(x_i) \, .
\end{align}
We can define the single-shot distribution $P_\varphi^s$ as the pair $P_\varphi^s = (P_\varphi(0), P_\varphi(1))$. The associated single-shot Fisher information is
\begin{align}
F_s:=F(P_\varphi^s) &= \sum_{x=0,1} {[\partial_\varphi P_\varphi (x)]^2 \over P_\varphi(x)} \nonumber \\ &= {4  \sin^2 (\varphi)} + {4 \cos^2 (\varphi) } \nonumber \\ &= 4
\end{align}
and since the Fisher information is additive for a product distribution we obtain
\begin{align}\label{f2}
F(P_\varphi) = \sum_{i=1}^N F(P^s_\varphi) = 4\sum_{i=1}^N =4N \, ,
\end{align}
in agreement with Eq.~(\ref{f}).

This standard quantum limit scaling follows from the use of the product distribution described in Eq.~(\ref{pd}). Sufficient conditions for obtaining a product distribution are that the $K_x$ are Hermitian and share an orthonormal eigenbasis $\{\ket{b_{1,2}}\}$, and that $\rho$ is one of the corresponding spectral projections, i.e., $\rho = \ket{b_1}\bra{b_1}$ or $\rho = \ket{b_2}\bra{b_2}$. In such a case, the resulting string of Kraus operators applied to the system commute and hence do not create any temporal correlations in the dynamics.  In the example above, Eqs.~(\ref{prop}) and (\ref{prop2}) imply that the $K_x$ are Hermitian and share a common orthonormal eigenbasis that may or may not depend on $\varphi$. We have in this case that
\begin{align}
K_x(\varphi) = \sum_{n=1,2} \lambda^n_x(\varphi) \ket{b_n}\bra{b_n} \, ,
\end{align}
where the eigenvalues $ \lambda^n_x(\varphi)$ depend in general on $\varphi$. If $\rho = \ket{b_1}\bra{b_1}$ say, then for $x=0,1$
\begin{align}
&P_\varphi(x) = {\rm tr}(K_x(\varphi)^2\rho) = \lambda_x^1(\varphi)^2,& \nonumber \\ &P_\varphi({\bf x}) = P_\varphi(0)^{k_x}P_\varphi(1)^{N-k_x} = \prod_{i=1}^N P_\varphi (x_i) \, ,
\end{align}
where $k_x$ is the number of $x_i=0$, and $N-k_x$ is the number of $x_i=1$, in the string ${\bf x}$. In order to get a $\varphi$-dependent result the eigenvalues $\lambda_x^n$ must depend on $\varphi$. Alternatively if the $(\lambda_x^n)^2$ are independent of $n$ as in the example from Eq.~(\ref{ch}) above, then $\rho$ can be a completely arbitrary density matrix and the same result will follow. In this case both $K_{0,1}^2$ are proportional to the identity $K_x^2 = \lambda_x^2 \, \mathbb{1}$, so that $P_\varphi(x) = {\rm tr}(K_x^2\rho)=\lambda_x^2$ for any normalised $\rho$.

\subsubsection{An example with enhanced precision}\label{example_enhanced}

\begin{figure}[t]
\centering
\includegraphics[width=0.48\textwidth]{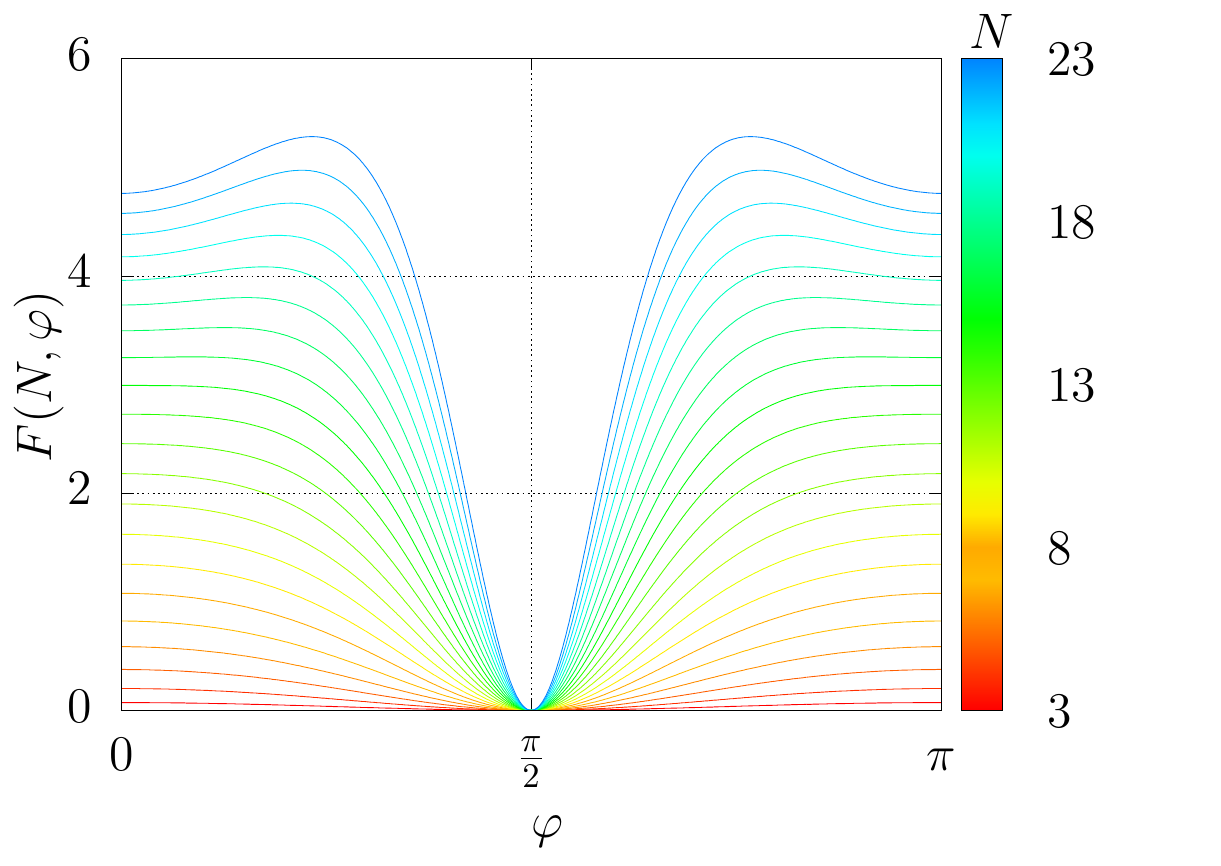}
\caption{[Colour online] Fisher information as a function of the parameter $\varphi$.  Each curve shows a different number of time steps $N$, as illustrated by the key.  Here the free parameters are chosen to be $A=0.9$ and $b=0.1$.  We see a steady increase in the Fisher information for all values of $\varphi$.  However, the peak value seems to move closer to $\pi/2$ as $N$ increases.}
\label{Fisherfigtimephase2}
\end{figure}

In the following, we consider a single qubit with parameter dependent resetting. More concretely, we reset into a state that is a function of a parameter $\varphi$ and whose measurements can be described by 
\begin{eqnarray} \label{aparameter}
K_0 & = & \left(\begin{array}{cc}
1 & 0 \\ 0 & A \end{array} \right)
\, , \nonumber \\
K_1 & = & \left( \begin{array}{cc}
0 & \cos(\varphi) \sqrt{1 - A^2} \\ 0 & \sin(\varphi) \sqrt{1 - A^2} \end{array} \right) 
\end{eqnarray}
with $0 \leq A \leq 1$.  Furthermore, we may choose an initial state given by the density matrix
\begin{eqnarray} \label{bparameter}
\rho & = & b \ket{0} \bra{0} + (1 - b) \ket{1} \bra{1} \, ,
\end{eqnarray}
where $0 \leq b \leq 1$. One can easily check that the above Kraus operators satisfy the completeness relation in Eq.~(\ref{Kraus-op}) but do not commute;
\begin{eqnarray} \label{bparameterxxx}
[K_0, K_1 ] &\neq & 0 \, .
\end{eqnarray}
Immediately, it should be noted that in general for this choice of Kraus operators, Eq.~(\ref{c1}) is different to Eq.~(\ref{c2}). Thus, these Kraus operators have potential to provide a Fisher information with greater than linear scaling.  In the next section, we will have a closer look at a possible implementation of the above $K_0$ and $K_1$.

\begin{figure}[t]
\centering
\includegraphics[width=0.48\textwidth]{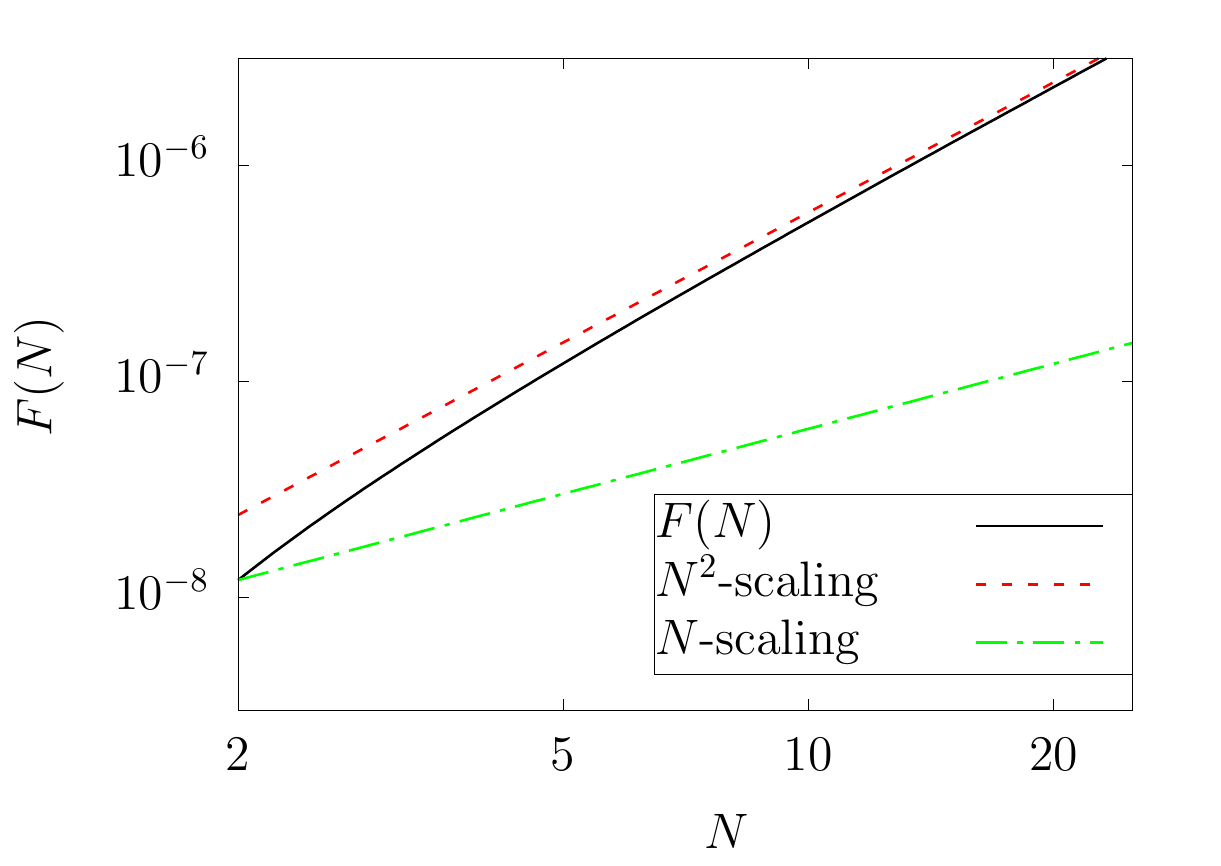}
\caption{Fitted log-log plot of the Fisher information for the Kraus operators given in Eq.~(\ref{Krausfinal}) for $\varphi = (499/500) \pi$ and with $A \approx 1$, such that $\sqrt{1-A^2} = 10^{-4}$.  The trend is clearly not linear and is therefore beyond the standard linear scaling of classical systems.  Plots showing scaling as $\sim N^2$ and $\sim N$ are shown for illustration.}
\label{Fisherrotate}
\end{figure}

\begin{figure}[t]
\centering
\includegraphics[width=0.48\textwidth]{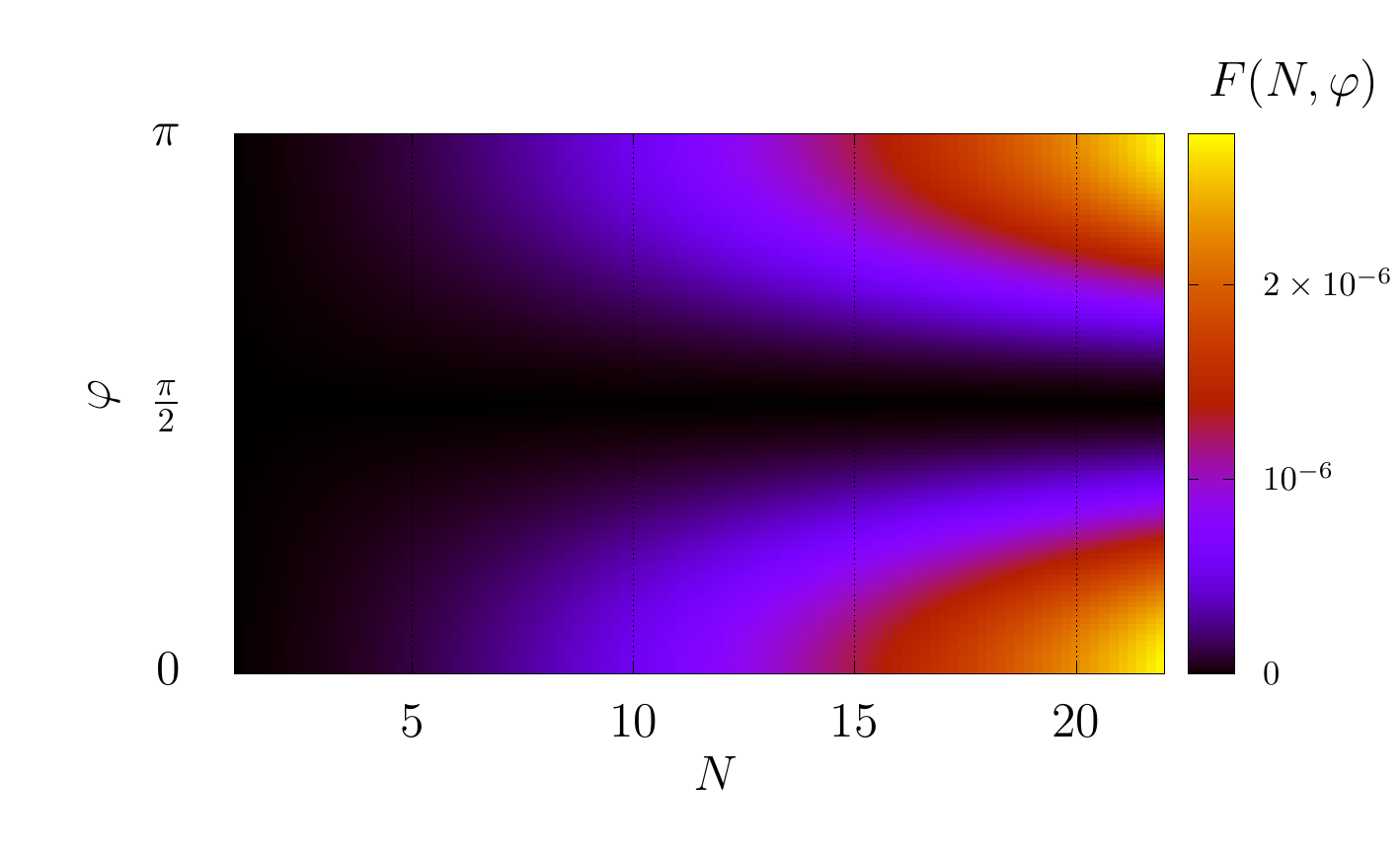}
\caption{[Colour online] Fitted function of the Fisher information for the scheme described by the Kraus operators in Eq.~(\ref{Krausfinal}), again taking $\sqrt{1-A^2} = 10^{-4}$.  We see that the Fisher information is maximised around $0$ and $\pi$ and also appears to grow non-linearly at these points, as shown in Fig.~\ref{Fisherrotate}.}
\label{Fisherfit}
\end{figure}

Fig.~\ref{Fisherfigtimephase2} shows the Fisher information generated by the above Kraus operators as a function of $\varphi$ and has been obtained from a numerical simulation of the sequential measurements and respective stochastic measurement sequences. In general, we observe a relatively complex Fisher information which does not follow a simple trend. Our numerical simulations show that the best measurement enhancement is achieved when the parameter $A$ in Eq.~(\ref{aparameter}) is close to unity, while the choice of $b$ offers little physical interest in long term scaling. In this case, we find that $F(N)$ scales as
\begin{eqnarray} \label{scale}
F(N) & \sim & \left(N^2 - N + c\right) \, , 
\end{eqnarray}
where $c$ is a small constant (c.f.~Figs.~\ref{Fisherrotate} and \ref{Fisherfit}). For large $N$, the right hand side of Eq.~(\ref{scale}) is dominated by the $N^2$ term which implies scaling approaching the Heisenberg limit. The numerical simulations moreover show that
\begin{eqnarray}
F(N,\varphi) & = & \cos^2(\varphi) \left(N^2 - N + c \right)
\end{eqnarray}
to a very good approximation. For example, Fig.~\ref{Fisherrotate} shows $F(N,\varphi)$ for a fixed value of $\varphi$. Fig.~\ref{Fisherfit}, which shows $F(N,\varphi)$ for $\varphi \in (0,\pi)$, clearly demonstrates a non-linear growth. 

In general, for most values of $A$, the Fisher information grows non-linearly with $N$ initially but assumes linear scaling in $N$ for large $N$. Only when the parameter $A$ is close to one, the Fisher information is non-linear for a relatively wide range of $N$. Our simulations show that enhanced scaling which overcomes the standard quantum limit can be generated with no entanglement. The exact behaviour that results from varying the parameter $A$ is not studied here, though presents a potentially interesting topic on how this affects the long-term scaling as a function of $T$.  For instance, we know that taking values of $A$ increasingly close to $1$ results in more persistant Heisenberg-scaling, but its asymptotic limit is not determined. In all the finite cases considered here the scaling eventually reaches that of the standard quantum limit, but still offers an enhancement for some $T$.  This result supports our earlier results in Ref.~\cite{PRA}, where we analyse a much more complex system with applications in quantum metrology. There is no reason why a quantum system occupying a larger Hilbert space should not persist with enhanced scaling further even for very large values of $N$. Unfortunately, numerical calculations of the Fisher information are in general difficult to obtain in these complex systems.  However, what we have seen here is that enhanced scaling is possible in such systems.

Moreover, we have seen in this section, that there are necessary (although not sufficient) conditions for producing a quantum enhancement.  Firstly, the Kraus operators should depend on $\varphi$ in a non-trivial way. Secondly, Kraus operators which correspond to different measurement outcomes should not commute with each other which is a necessary criteria for strong temporal correlations in the measurement statistics. This observation can be used to guide the design of quantum jump metrology schemes which can then be analysed for instance numerically via the simulation of the proposed measurement scheme.

\section{Detailed analysis of a concrete physical implementation} \label{Toy}

To obtain more insight into how this works, we finally have a closer look at a possible concrete implementation of Eq.~(\ref{aparameter}) and for which we can obtain analytical results. Suppose a single two-level atom is allowed to freely decay, while being subjected to parameter-dependent back-action upon photon emission. Such parameter-dependent queries could be realised physically by connecting the photodetectors that monitor the radiation field around the atom to a laser directed towards the atom. The laser sends a feedback pulse to the atom whenever a photon is detected. Most importantly, the laser parameters of the feedback pulse should depend on the unknown parameter $\varphi$ that we want to measure. In this case only the Kraus operator $K_1$ would be $\varphi$-dependent. There are a number of ways such a scheme could be implemented but we focus on the resulting behaviour here.

Proceeding as described for example in Refs.~\cite{Reset,Molmer,Carmichael,HQMM} and as we shall see below, we find that the dynamics of such a two-level atom with ground state $|0 \rangle$ and an excited state $|1 \rangle$ over a short time interval $\Delta t $ can be described by the Kraus operators
\begin{eqnarray} \label{Krausfinal}
K_0 & = & \ket{0} \bra{0} + {\rm e}^{- {1 \over 2} \Gamma \Delta t} \ket{1} \bra{1} \nonumber \\
K_1 & = & \sqrt{\Gamma \Delta t} \, \left[ \cos(\varphi) \ket{0} \bra{1} - {\rm i} \sin(\varphi) \ket{1} \bra{1} \right] 
\end{eqnarray}
Here $\Gamma$ denotes the spontaneous photon emission rate of the atom and it is assumed that every detection of a photon results in a rotation of the atomic state by an angle $\varphi$. In the next subsection we see that the quantum trajectories of the atom are due to the successive application of the above Kraus operators on a coarse grained time scale $\Delta t$. To measure the unknown parameter $\varphi$, we observe the average number $\bar N(T,\varphi)$ of photons emitted by the atom in a time interval $(0,T)$ of length $T$ which we derive later in this section. Eventually, we show that this quantity may provide an enhanced measurement of the unknown parameter $\varphi$.

Notice also that the Kraus operators $K_0$ and $K_1$ in Eq.~(\ref{Krausfinal}) coincide with the Kraus operators in Eq.~(\ref{aparameter}) for $A={\rm e}^{- {1 \over 2} \Gamma \Delta t} $ in the limit of frequent measurements on the free radiation field which implies small time intervals $\Delta t$. The Kraus operators $K_0$ and $K_1$ in Eq.~(\ref{Krausfinal}) are in fact the Kraus operators used in Figs.~\ref{Fisherrotate} and \ref{Fisherfit} to allow for a comparison with the results in this section.  

\subsection{Quantum jump operators in open quantum systems}

From quantum optics, we know that an atom that is constantly monitored but does not emit a photon evolves with the conditional Hamiltonian \cite{Reset,Molmer,Carmichael}
\begin{eqnarray}
H_{\rm cond} &=&  -\frac{{\rm i}}{2} \hbar \Gamma \sigma^+ \sigma^- \, ,
\end{eqnarray}
which is non-Hermitian. If no photon is detected for a short time $\Delta t$, the state of the atom evolves into the unnormalised state 
\begin{eqnarray} \label{condtime}
|\psi_{\rm I}(t+\Delta t) \rangle &=& \exp \left(- {\rm i} H_{\rm cond} \Delta t/\hbar \right) \, |\psi_{\rm I}(t) \rangle
\end{eqnarray}
up to first order in $\Delta t$. The normalisation of this state squared equals the probability for no photon emission in $(t,t+\Delta t)$. Hence, the no-photon time evolution of the atom automatically implements the transformation $|\psi_{\rm I} \rangle \longrightarrow K_0 \, |\psi_{\rm I} \rangle$ with $K_0$ given in Eq.~(\ref{Krausfinal}), as long as $\Delta t $ is sufficiently small. This can be shown by calculating the right hand side of Eq.~(\ref{condtime}).

Whenever a photon is detected, the atom is subsequently found in its ground state $|0 \rangle$. Moreover we know that the probability density for the emission of a photon is the product of its spontaneous decay rate $\Gamma$ and the population $\|\langle 1|\psi (t) \rangle\|^2 $ in the excited state. Suppose now, every photon emission triggers a short strong laser pulse which transfers its state into a state of the form
\begin{eqnarray} \label{reset}
\ket{\psi^{{\rm ph}}} & = & \cos (\varphi) \ket {0} - {\rm i} \sin (\varphi) \ket{1} 
\end{eqnarray}
which could be achieved in a variety of ways. Then the change of atomic state in the case of an emission can be described by the Kraus operator $K_1$ in Eq.~(\ref{Krausfinal}). 

\subsection{Average number of emitted photons}

In order to determine the unknown parameter $\varphi$, we utilise in the following a measurement of the average number of emitted photons in a time period $(0,T)$, denoted $\bar N(T,\varphi)$. In this subsection, we calculate this observable for the proposed experimental setup.  To do so, we notice that $\bar N(T,\varphi)$ can be written as
\begin{eqnarray} \label{Nbar}
\bar{N}(T,\varphi) & = & \sum\limits_{n=1}^\infty n p_n(0,T) \, ,
\end{eqnarray}
where $p_n(0,T)$ is the probability of the system emitting exactly $n$ photons in a time interval $(0,T)$ for a given initial state. For simplicity, we assume in the following that the state of the atom at the time $t=0$ equals the the reset state after a photon detection $\ket{\psi^{{\rm ph}}}$ which can be found in Eq.~(\ref{reset}). 

Next we notice that the time evolution operator of our two-level system under the condition of no photon detections equals $U_{\rm cond} (T, 0) = {\rm exp}\left(- {\rm i} H_{\rm cond} T /\hbar \right)$ and that the probability of the system not emitting a photon in a time period $(0,T)$, $p_0(0,T)$ is given by
\begin{eqnarray} \label{reset-state}
p_0(0,T) & = & \| U_{\rm cond} (T, 0) \, \ket{\psi^{{\rm ph}}} \|^2 \nonumber \\
&=& \cos^2(\varphi) + {\rm e}^{- \Gamma T} \sin^2(\varphi) \, .
\end{eqnarray}
Moreover, one can show that the probability density for emitting exactly one photon in a time period $(0,T)$ at a time $t$ equals the probability density $w_1(0,t)$ for the emission of a first photon at a time $t$,
\begin{eqnarray} \label{w1}
w_1(0,t) & = & - \frac{{\rm d}}{{\rm d}t} p_0 (0,t) = \Gamma \sin^2 (\varphi) {\rm e}^{-\Gamma t} \, ,
\end{eqnarray}
multiplied by the probability for no photon emission in $(t,T)$ which we denote $ p_0(t,T)$. The probability $p_1(0,T)$ of the system emitting exactly one photon in a time period $(0,T)$ is hence obtained by integrating these probability densities over $t$. Hence
\begin{eqnarray} \label{p10T}
p_1(0,T) & = & \int\limits_0^T {\rm d} t \, w_1(0,t) p_0(t,T) \, .
\end{eqnarray}
Proceeding analogously and calculating the probability for exactly $n$ photon emissions in a time interval $(0,T)$ moreover yields
\begin{eqnarray} \label{iter}
p_n(0,T) & = & \int\limits_{0}^T {\rm d} t_1 w_1(0,t_1) p_{n-1}(t_1,T) \, ,
\end{eqnarray}
where $p_{n-1}(t_1,T)$ denotes the probability for the emission of $n-1$ photon in the time interval $(t_1,T)$. In other words, the probability for having $n$ photons in $(0,T)$ is the sum of all probability densities with a first photon at $t_1 \in (0,T)$ and exactly $n-1$ photons in $(t_1,T)$. In the following, we use this relation to determine $p_n$ as a function of $w_1$ and $p_0$.
  
Iteration of Eq.~(\ref{iter}) yields 
\begin{eqnarray}
p_n (0,T) & = & \int\limits_{0}^T {\rm d} t_1 \int\limits_{t_1}^T {\rm d} t_2 w_1(0,t_1) w_1(t_1,t_2) p_{n-2}(t_2,T) \nonumber \\
\end{eqnarray}
and so on. Hence
\begin{eqnarray}
p_n (0, T) & = & \int\limits_{0}^T {\rm d}t_1 \int\limits_{t_1}^T {\rm d} t_2 \dots \int\limits_{t_{n-1}}^T {\rm d} t_n \, w_1(0,t_1) \nonumber \\
&& \times  w_1(t_1,t_2) \dots w_1(t_{n-1},t_n) p_0(t_n,T) \nonumber \\
& = & \prod_{i=1}^n \left( \, \, \, \int\limits_{t_{i-1}}^T {\rm d} t_i \, w_1 (t_{i-1},t_i) \right) p_0(t_n , T) ~~~
\end{eqnarray}
with $t_0 = 0$. If we consider the case where we wait for a large amount of time such that we may take the stationary limit $T \rightarrow \infty$, we find these integrals factorise nicely, meaning
\begin{eqnarray}
\lim_{T \rightarrow \infty} \int\limits_{t_{i-1}}^T {\rm d} t_i \, w(t_{i-1},t_i) & = & \sin^2(\varphi) \, , \nonumber \\
\lim_{T \rightarrow \infty} p_0(t_n,T) & = & \cos^2(\varphi) \, .
\end{eqnarray}
Hence, we find that the probability for $n$ photons in the stationary limit is given by
\begin{eqnarray} \label{pninfinity}
\lim_{T \rightarrow \infty} p_n(0,T) & = & \sin^{2n} (\varphi) \cos^2 (\varphi)
\end{eqnarray}
The average number of photons emitted in the stationary limit can now be calculated by substituting Eq.~(\ref{pninfinity}) into Eq.~(\ref{Nbar}), which gives
\begin{eqnarray}
\bar{N} (\infty,\varphi) & = & \sum\limits_{n=1}^\infty n \sin^{2n} (\varphi) \cos^2 (\varphi) \, .
\end{eqnarray}
This is nearly a geometric series.  After appropriately modifying the standard geometric series, it can be shown that
\begin{eqnarray}
\sum\limits_{n=1}^\infty n r^n & = & \frac{r}{\left(1-r\right)^2} \, .
\end{eqnarray}
Taking $r = \sin^2(\varphi)$, we hence find
\begin{eqnarray} \label{Nbarlimit}
\bar{N} (\infty,\varphi) & = & \tan^2(\varphi) \, .
\end{eqnarray}
This function matches expectations, as we see that for the case where the system is reset exactly to the excited state, we see an infinite number of photons, whereas when it is reset to the ground state we see no photons.

\begin{figure}[t]
\centering
\includegraphics[width=0.48\textwidth]{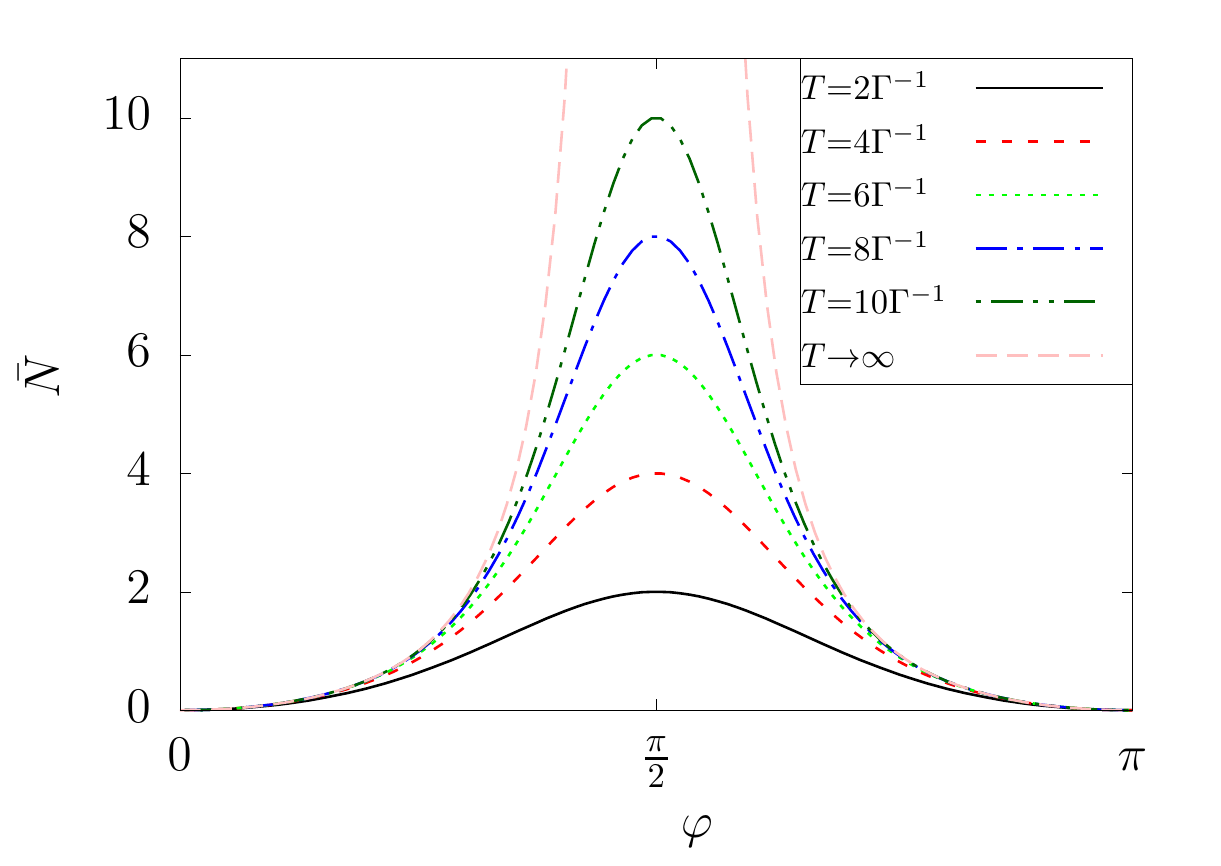}
\caption{The plot of $\bar{N}$ now does not go to infinity, as a finite amount of time is considered.  The curve has a similar functional shape to the case of infinite time and hence demonstrates the validity of the calculations.  Here, the sum is taken up to $n=2000$.  The limit of $T \rightarrow \infty$ calculated in Eq.~(\ref{Nbarlimit}) is also shown for consistency.}
\label{Two-level-Nbar-var-phi}
\end{figure}

For the purposes of metrology, we want a signal we can scale with time.  As such, we can calculate how this signal scales for finite $T$.  By not imposing $T \rightarrow \infty$, the integrals no longer factorise nicely.  Nevertheless, a solution can still be found for $p(n,T)$, which is given by
\begin{eqnarray} \label{pn0T}
p_n(0,T) & = & \sin^{2n} (\varphi) \cos^2 (\varphi) + \frac{{\rm e}^{-\Gamma T} \sin^{2n}(\varphi)}{n!} \nonumber \\
&& \times \left(\left(\Gamma T\right)^n - \cos^2(\varphi) \sum\limits_{m=0}^n \frac{n!}{m!} \left(\Gamma T\right)^m \right) \, . \nonumber \\
\end{eqnarray}
The derivation of Eq.~(\ref{pn0T}) is given in App.~\ref{AppA}.  The limit of the sum in Eq.~(\ref{Nbar}) where $n \rightarrow \infty$ is now more difficult to resolve.  Although the limit is well defined, it is not straightforward to explicitly calculate.  Hence, for simplicity, all results involving this term will be approximated by choosing a large finite value for $n$.  In doing so, $\bar{N} (T,\varphi)$ can be calculated to a very good approximation.  In Fig.~\ref{Two-level-Nbar-var-phi}, we see how this function behaves as a function of $\varphi$ at a variety of times $T$.

\begin{figure}[t]
\centering
\includegraphics[width=0.48\textwidth]{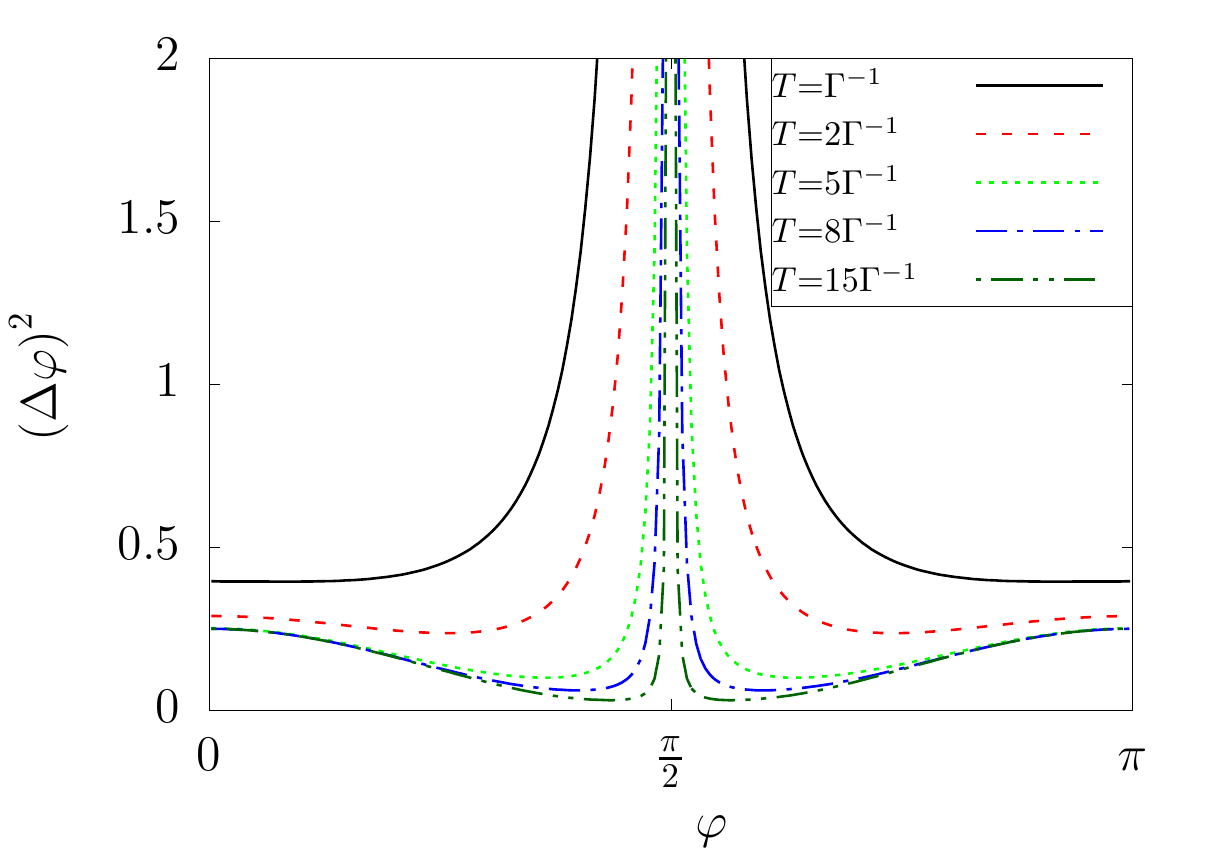}
\caption{Uncertainty $\left(\Delta \varphi\right)^2$ plotted as a function of $\varphi$.  Initially, the uncertainty is minimised at multiples of $\pi$.  However, these uncertainties do not significantly decrease in time.  As $T$ increases, the optimum value of $\varphi$ for measurement moves closer to $\pi/2$. This result is again produced with a sum up to $n=2000$.}
\label{Two-level-var-phi}
\end{figure}

This signal clearly displays dependence on the parameter $\varphi$ that grows in time.  Hence it should be possible to use this signal to extract information about $\varphi$.  In order to calculate the uncertainty in $\varphi$, we use the error propagation formula \cite{review}
\begin{eqnarray} \label{Error}
\left( \Delta \varphi \right)^2 & = & \frac{\left(\Delta A(\varphi) \right)^2}{\left| \frac{{\rm d} A}{{\rm d}\varphi}\right|^2} \, ,
\end{eqnarray}
for some signal $A(\varphi)$ that has dependence on the unknown parameter $\varphi$.  For our case of $A = \bar{N}$, the variance in the numerator is given by
\begin{eqnarray}
\left( \Delta \bar{N} \right)^2 & = & \sum\limits_{n=1}^\infty n^2 p_n(0,T) - \left( \sum\limits_{n=1}^\infty n p_n(0,T) \right)^2 ,
\end{eqnarray}
from Eq.~(\ref{Nbar}), while the derivative in the denominator can be calculated straightforwardly.

\begin{figure}[t]
\centering
\includegraphics[width=0.48\textwidth]{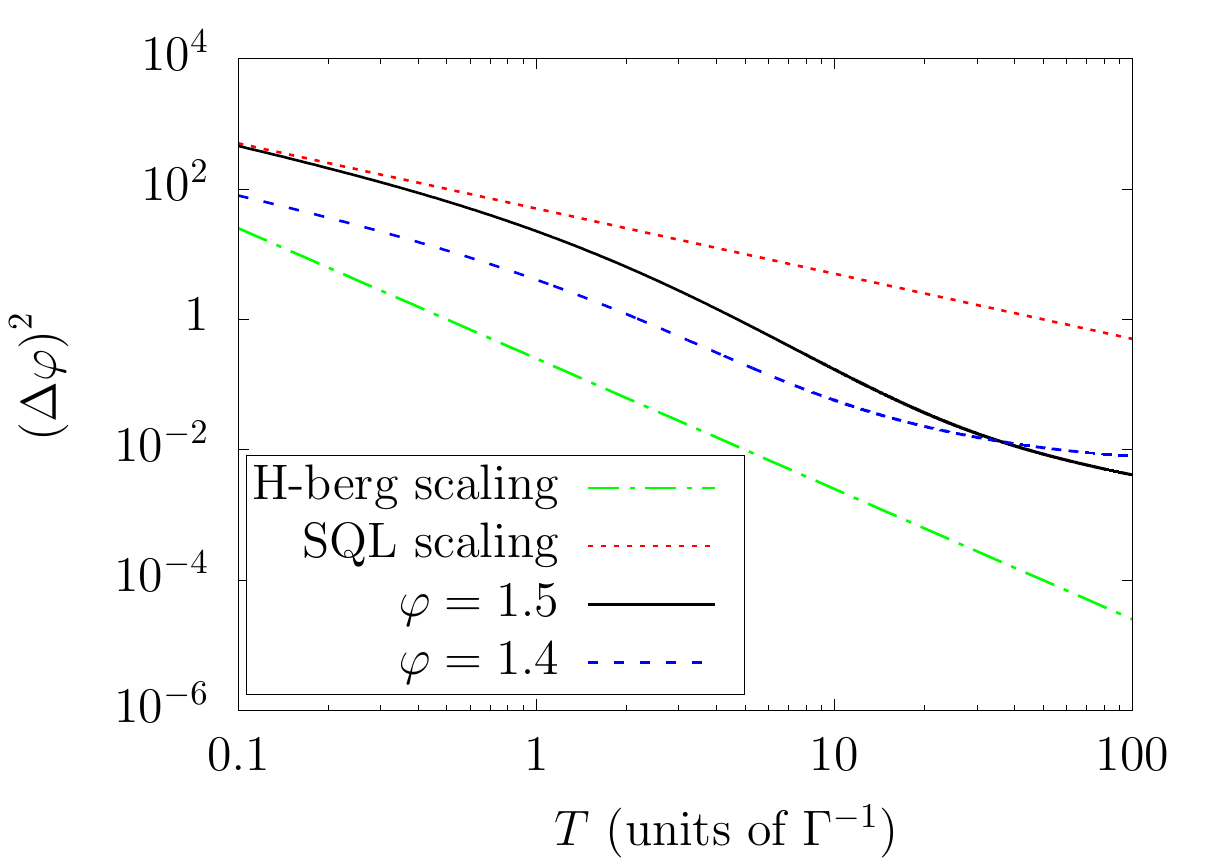}
\caption{The uncertainty $\left(\Delta \varphi\right)^2$ as a function of $T$ plotted for fixed values of $\varphi$ ($\varphi = 1.4,1.5$). For illustrative purposes, scaling according to the standard quantum limit ($\sim 1/T$) and the Heisenberg limit ($\sim 1/T^2$) are shown.  We see that the scaling of our system lies between these two.  The results are produced with a sum up to $n=2000$ again.}
\label{Two-level-var-T}
\end{figure}

Plotting as a function of $\varphi$ for a variety of times $T$, we see in Fig.~\ref{Two-level-var-phi} how the uncertainty in $\varphi$ changes in time.  In particular, the error decreases in time.  However, it appears to approach a fixed point that depends on the value of $\varphi$ being considered.  Also, the error is able to reach a lower value for a large amount of time the closer it is to $\pi/2$.  Hence, to maximise the scaling it appears that we should choose a value of $\varphi$ close to $\pi/2$, so long as a large time $T$ may be considered.  As such, when $T$ is unable to be taken as a large value, a value of $\varphi$ away from $\pi/2$ is preferable.  Taking $\varphi$ as values close to $\pi/2$, we now plot $\left(\Delta \varphi\right)^2$ as a function of time $T$.  This is shown in Fig.~\ref{Two-level-var-T}.  Here, we see the scaling is surpassing that of the standard quantum limit.

Crucially, we see that there is an enhanced scaling present for this measurement scheme.  Although this measurement is not necessarily an optimum measurement, it serves as a proof-of-principle that an enhanced time-dependent scaling can be found for a relatively simple system with quantum feedback.  Indeed, there are many ways in which this system can be developed further, including going to a larger system size or performing a more complex measurement, such as using photon correlations where an enhancement has already been shown \cite{PRA}.  In Fig.~\ref{Two-level-var-T}, we see that the uncertainty in $\varphi$ seems to be levelling off to a fixed value.  This is also suggested in Fig.~\ref{Two-level-var-phi} for other values of $\varphi$.  If we move to a larger system size, the overall uncertainty should be reduced further.  This is because in a larger system size two initially close together points in the relevant space can move further away from each other and hence become more distinguishable.

\begin{figure}[t]
	\centering
	\includegraphics[width=0.48\textwidth]{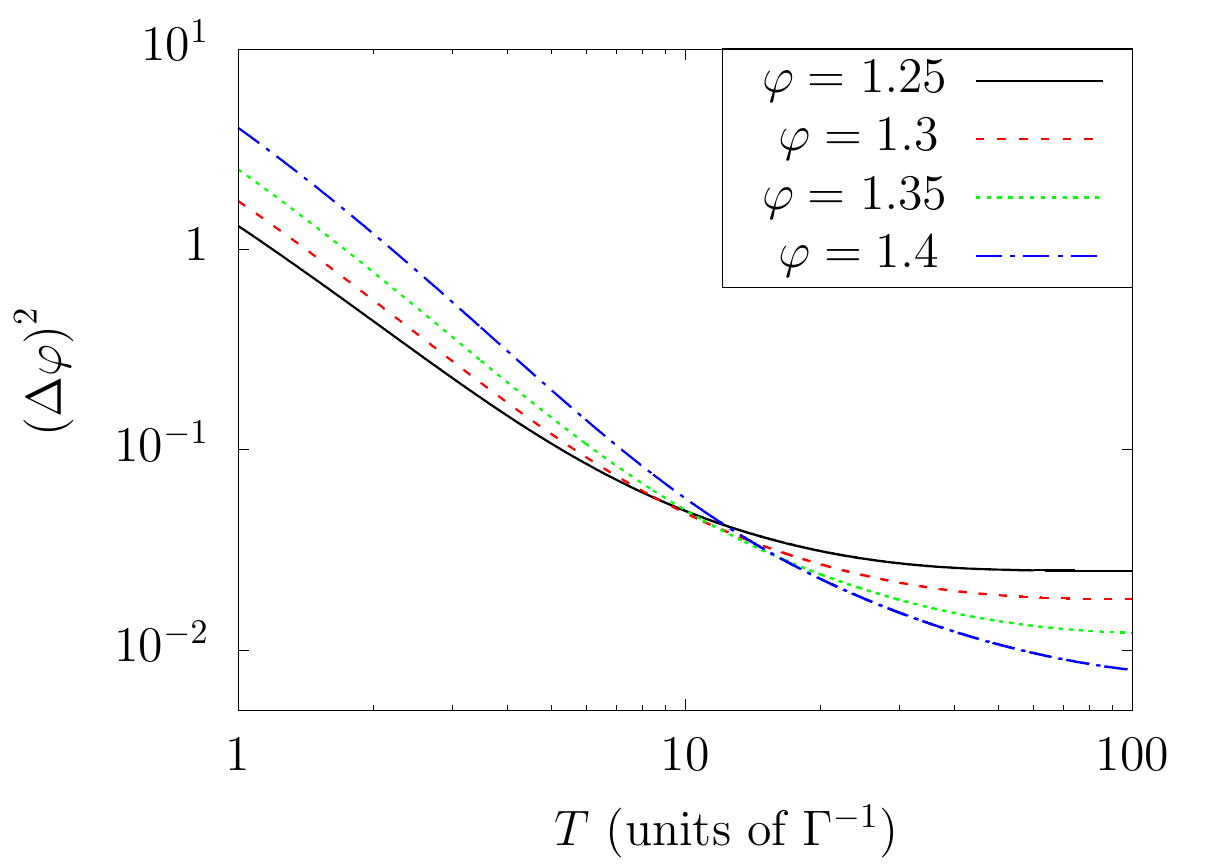}
	\caption{Uncertainty $\left(\Delta \varphi\right)^2$ as a function of $T$ plotted for a range of values of $\varphi$.  For smaller values of $\varphi$, we see the long time limit of plateauing behaviour both more clearly and earlier.  This shows how values of $\varphi$ that are initially optimal in the short time limit eventually become inferior to others in the long time limit.  The results are produced with the same simulations as previous plots in this section.}
	\label{Two-level-var-T-plateau}
\end{figure}

Another observation from Figs.~\ref{Two-level-var-phi} and \ref{Two-level-var-T} is that for small $T$ a value of $\varphi$ closer to $0$ is preferable.  However, as $T$ increases, the optimum value of $\varphi$ to be measured shifts asymptotically closer to $\pi/2$. This supports what was seen earlier in Fig.~\ref{Fisherfigtimephase2}, where the maximum of the Fisher information moves closer to $\pi/2$ with increasing $N$.  Indeed all values of $\varphi$ tend to follow a standard quantum limit scaling initially for this measurement scheme, before eventually at some time $T$ gaining some enhanced scaling and ultimately then plateauing at some fixed value for $\left(\Delta \varphi \right)^2$.  This allows one to determine an optimum value of $\varphi$ to measure if $T$ becomes a limiting factor.  This is shown for a range of values of $\varphi$ in Fig.~\ref{Two-level-var-T-plateau}.

\section{Conclusions} \label{Conclusions}

This paper introduces the general concept of {\em quantum jump metrology} which is based on generalised sequential measurements and considers the total duration $T$ of the measurement process as the main measurement resource. One way of implementing quantum jump metrology is to apply quantum feedback to open quantum systems. It is shown that this approach can indeed result in precision scaling beyond the standard quantum limit without the need for complex state preparation. This is in contrast to closed quantum systems, where overcoming the standard quantum limit requires entanglement or the presence of other highly-non-classical states which are hard to prepare experimentally. Open quantum systems therefore currently receive a lot of attention in quantum metrology but their systematic study is often difficult, since standard quantum metrology techniques do not extend easily to more complex systems \cite{review2}.

Here we provide novel insight into quantum metrology with generalised sequential measurements by drawing analogies to Hidden Quantum Markov Models \cite{Wiesner,Monras,Schuld,Biamonte,Bristol}. This analogy suggests that there could be a wide range of computational advantages compared to analogous classical machines \cite{Monras,Yeh,indian,Crutchfield}.  As usual, we describe the system dynamics induced by the generalised measurements with Kraus operators. For applications in quantum metrology, these should depend in a non-trivial fashion on the parameter $\varphi$ that we want to measure. Moreover, the Kraus operators associated with different measurement outcomes should not commute with each other. A quantum enhancement of the scaling of errors can be expected, when measurement sequences cannot be modelled as Markov processes and contain long-range temporal correlations. The above described necessary (although not sufficient) conditions can be used to guide the design of quantum metrology schemes in open quantum systems.

To substantiate our claims and to demonstrate the practicality of our approach, Section \ref{Toy} finally analyses a  quantum metrology scheme which consists only of a two-level atom with spontaneous photon emission and external laser driving. Due to its simplicity, it is possible to analyse the precision of parameter estimates deduced from the atomic dynamics with relative ease. In doing so, we show that observing the output of an open quantum system with sequential measurements and quantum feedback can indeed be used to exceed the standard quantum limit. This result is consistent with our earlier numerical analysis of a more complex quantum metrology scheme based on the conditional dynamics of the coherent states of an optical cavity for which we were unable to establish analytical bounds for the precision of measurement outcomes \cite{PRA}.

We also emphasise again here that the approach and implentation shown are not necessarily optimum, but provide a simple pathway to enhancements.  There are other methods that obtain enhancements that could potentially also be incorporated into this work, such as a final measurement of the system to obtain a further enhancement \cite{Kiilerich}.  The scope for further developments of these schemes is large and should be of significant interest.  Furthermore, as discussed in Sec.~\ref{example_enhanced}, there is still much more to explore in terms of the Fisher information of systems of this type, such as a more rigorous study involving varying the parameter $A$.  We leave this as an open question here that remains to be investiagted in future work.  Overall, we hope that the general discussion of this paper helps the design of novel practical quantum metrology schemes. \\[0.5cm]

\noindent {\em Acknowledgement.} We acknowledge financial support from the Oxford Quantum Technology Hub NQIT (grant number EP/M013243/1). Statement of compliance with EPSRC policy framework on research data: This publication is theoretical work that does not require supporting research data.

\appendix
\section{Derivation of $p_n(0,T)$} \label{AppA}
	We derive Eq.~(\ref{pn0T}) by direct construction. We start with the expression
	\begin{align} \label{pn+10Tdef}
	p_{n+1}(0,T) = \int_0^T d t\, w_1(0,t) p_n(t,T).
	\end{align}
	Using the notation $p_n(0,t)=: p_n(t)$ and $w_1(0,t)=: w_1(t)$ and noting that $p_n(t,T)=p_n(0,T-t)=: p_n(T-t)$ the convolution theorem yields
	\begin{align}
	{\tilde p}_{n+1}(s)={\tilde w}_1(s) {\tilde p}_n(s) = {\tilde w}_1(s)^{n+1} {\tilde p}_0(s)
	\end{align}
	where ${\tilde g}:={\mathcal L}[g]$ denotes the Laplace transform of $g$. Applying the convolution theorem again then yields
	\begin{align}\label{pnp1}
	p_{n+1}(T)=(f_{n+1} * p_0)(T),\qquad f_n(t) = {\mathcal L}^{-1}[{\tilde w}_1(s)^n](t)
	\end{align}
	where $*$ denotes the convolution product;
	\begin{align}
	(f*g)(T) := \int_0^T dt\, f(t)g(T-t).
	\end{align}
	Using Eq.~(\ref{w1}) we have
	\begin{align}
	{\tilde w}_1(s)^{n+1} &= {\Gamma^{n+1}\sin^{2(n+1)}(\varphi) \over (\Gamma+s)^{n+1}} \nonumber \\ &=  {\Gamma^{n+1}\sin^{2(n+1)}(\varphi) \over (-1)^n n! }{d^n\over ds^n}{1\over s+\Gamma} \nonumber \\ &= {\mathcal L}\left[\Gamma^{n+1}\sin^{2(n+1)}(\varphi) {t^n e^{-\Gamma t} \over n!} \right](s)
	\end{align}
	from which it follows that
	\begin{align}
	f_{n+1}(t) = \Gamma^{n+1}\sin^{2(n+1)}(\varphi) {t^n e^{-\Gamma t} \over n!}.
	\end{align}
	Using this expression and Eq.~(\ref{pnp1}) we obtain
	\begin{align}\label{pnp1b}
	p_{n+1}(T) =&  {\Gamma^{n+1}\sin^{2(n+1)}(\varphi) \over n!}\bigg[{e^{-\Gamma T} T^{n+1} \sin^2(\varphi) \over n+1} \nonumber \\ &\qquad \qquad +\cos^2(\varphi)\int_0^T dt\,  t^n e^{-\Gamma t}\bigg].
	\end{align}
	The integral in the above can be evaluated as
	\begin{align}\label{intgam}
	\int_0^T dt\, t^ne^{-\Gamma t} = {1\over \Gamma^{n+1}}\left[n!-\gamma(n+1,\Gamma T)\right]
	\end{align}
	where $\gamma$ denotes a special function called the incomplete $\gamma$-function, which is defined by
	\begin{align}\label{intgam2}
	\gamma(a,y)=\int_y^\infty dz\, z^{a-1}e^{-z}.
	\end{align}
	Using Eq. (\ref{intgam}) we obtain
	\begin{widetext}
	\begin{align}\label{pnp12}
	p_{n+1}(t) =& \sin^{2(n+1)}(\varphi)\cos(\varphi) + {(\Gamma T)^{n+1}e^{-\Gamma T}\sin^2(\varphi) \sin^{2(n+1)}(\varphi) \over (n+1)!}  -{\cos^2(\varphi) \sin^{2(n+1)}(\varphi) \over n!}\gamma(n+1,\Gamma T) \nonumber\\ =& \sin^{2(n+1)}(\varphi)\cos(\varphi) + {(\Gamma T)^{n+1}e^{-\Gamma T}\sin^{2(n+1)}(\varphi) \over (n+1)!} -\cos^2(\varphi) \sin^{2(n+1)}(\varphi) \times\left[{(\Gamma T)^{n+1}e^{-\Gamma T}\over (n+1)!} + {\gamma(n+1,\Gamma T)\over n!}\right]
	\end{align}
	\end{widetext}
	Using the definition (\ref{intgam2}) and integration by parts one can prove inductively that
	\begin{align}
	\gamma(n+1,\Gamma T) = e^{-\Gamma T} \sum_{m=0}^n {n!\over m!}(\Gamma T)^m.
	\end{align}
	We therefore obtain
	\begin{align}\label{pnp13}
	p_{n+1}(t)=& \sin^{2(n+1)}(\varphi)\cos(\varphi) \nonumber \\ &+ {(\Gamma T)^{n+1}e^{-\Gamma T}\sin^{2(n+1)}(\varphi) \over (n+1)!}\nonumber \\ &-\cos^2(\varphi) \sin^{2(n+1)}(\varphi)\sum_{m=0}^{n+1} {1\over m!}(\Gamma T)^m,
	\end{align}
	which is equivalent to the expression shown in Eq.~(\ref{pn0T}).

\end{document}